\documentclass[reprint,showpacs,preprintnumbers,amsmath,amssymb, aip]{revtex4-1}
\usepackage{graphicx,wasysym}% Include figure files
\usepackage{dcolumn}% Align table columns on decimal point
\usepackage{bm}% bold math
\usepackage{natmove}
\usepackage{tabulary}

\begin{document}

\preprint{1}

\title{Concentration Fluctuations and Capacitive Response in Dense Ionic Solutions}
\author{Betul Uralcan}

\affiliation{%
Department of Chemical and Biological Engineering, Princeton University, Princeton NJ 08544, United States
}%

\author{Ilhan A. Aksay}
\affiliation{%
Department of Chemical and Biological Engineering, Princeton University, Princeton NJ 08544, United States
}%

\author{Pablo G. Debenedetti}
\affiliation{%
Department of Chemical and Biological Engineering, Princeton University, Princeton NJ 08544, United States
}%

\author{David T. Limmer}
 \email{dlimmer@princeton.edu}
\affiliation{%
Princeton Center for Theoretical Science, Princeton University, Princeton NJ 08544, United States
}%

\date{\today}

\begin{abstract}
\subsection*{Abstract}
We use molecular dynamics simulations in a constant potential ensemble to study the effects of solution composition on the electrochemical response of a double layer capacitor. We find that the capacitance first increases with ion concentration following its expected ideal solution behavior, but decreases upon approaching a pure ionic liquid in agreement with recent experimental observations. The non-monotonic behavior of the capacitance as a function of ion concentration results from the competition between the independent motion of solvated ions in the dilute regime and solvation fluctuations in the concentrated regime. Mirroring the capacitance, we find that the characteristic decay length of charge density correlations away from the electrode is also non-monotonic. The correlation length first decreases with ion concentration as a result of better electrostatic screening but increases with ion concentration as a result of enhanced steric interactions. When charge fluctuations induced by correlated ion-solvent fluctuations are large relative to those induced by the pure ionic liquid, such capacitive behavior is expected to be generic.
\end{abstract}
\pacs{}% PACS, the Physics and Astronomy
                             % Classification Scheme.

\maketitle

Recent experimental observations have shown that the differential capacitance of a room temperature ionic liquid based electrical double layer capacitor can change markedly with solvent concentration \cite{bozym2015anomalous,liu2012effects,siinor2012influence}. Using molecular dynamics (MD) simulations we show that the concentration dependence of the capacitance results from the interplay between two different limiting behaviors. In the dilute ion concentration regime, charge fluctuations are simply proportional to the number of ions near the electrode, because their mean distances are larger than the electrostatic screening length and so those fluctuations are uncorrelated \cite{fedorov2014ionic}. In the pure ionic liquid, ions are densely packed and charge fluctuations are determined by steric constraints and interionic Coulomb correlations \cite{kornyshev2007double, fedorov2008towards, fedorov2008ionic, fedorov2008err, bazant2011double, bazant2011err, wu2011classical}. The addition of a small amount of solvent mediates these constraints, increasing the magnitude of charge fluctuations. At a specific concentration these effects are balanced, leading to an intermediate concentration where the capacitance is maximized. The analysis presented here offers a general way to understand the molecular contributions to the electrochemical response of complex electrolyte solutions, opening new directions for the optimization and rational design of energy storage devices.

Due to the increasing interest in ionic liquid based capacitors that provide high energy density devices \cite{simon2008materials,miller2008electrochemical}, ionic liquid-metal interfaces have become a recent focus of research \cite{mezger2015solid,mezger2009layering,hayes2011double,atkin2011situ,carstens2012situ,fedorov2014ionic,kornyshev2007double,merlet2014electric,limmer2015interfacial}. While the limiting behaviors of neat ionic liquids or their dilute solutions have been widely studied, relatively little is known about the properties of concentrated electrolytes at charged solid interfaces. Concentrated ionic solutions are practically relevant as they have the potential to exhibit both high energy and high power densities \cite{arulepp2004influence}. Experimentally, we have recently shown that the effect of solvent concentration on the differential capacitance of 1-Ethyl-3-methylimidazolium bis(trifluoromethylsulfonyl)imide ([EMIm$^+$][TFSI$^-$]) on glassy carbon electrodes that possess a large space charge capacitance \cite{Pope2015Four} is significant, with the capacitance increasing with the addition of a small amount of solvent\cite{bozym2015anomalous}. The capacitance decreases with further dilution resulting in an anomalous capacitance maximum at intermediate ion concentrations. Previous computational studies have suggested that the addition of a solvent to an ionic liquid has a small effect on differential capacitance \cite{feng2011counter,li2014interfaces}, though Merlet et al. have shown that solvent addition suppresses ionic overscreening at intermediate concentrations \cite{merlet2013influence}. 
%Our previous work has suggested that solvent fluctuations play a key role in the occurrence of the capacitance maximum\cite{bozym2015anomalous}. At present, a molecular understanding of the concentration dependence of the capacitance across a wide range of compositions is lacking.

\begin{figure*}[t]
\begin{center}
\includegraphics[width=15cm]{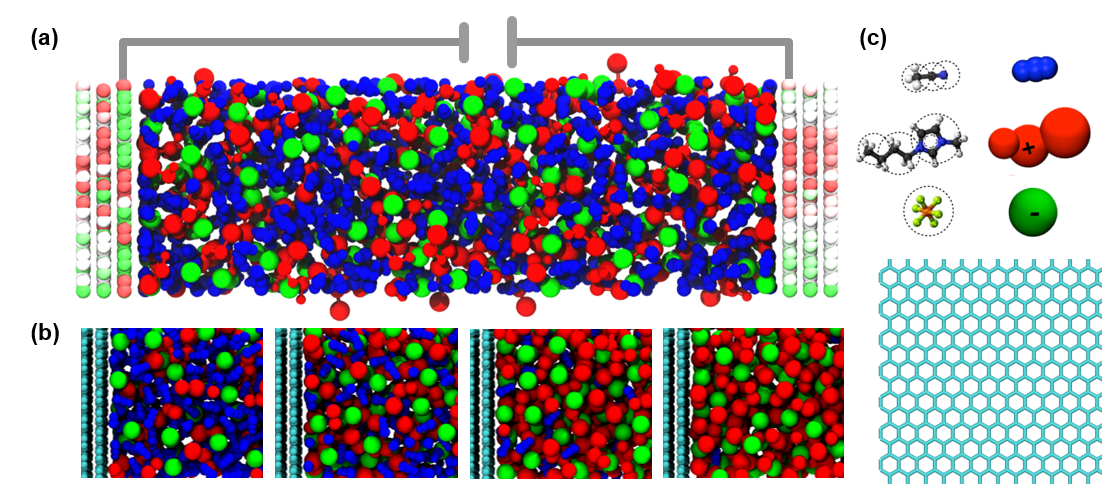}
\caption{Simulated capacitor. (a) Each capacitor consists of an electrolyte between two electrodes maintained at constant potential difference. The color code on the electrode atoms indicates the instantaneous charge on each carbon atom. (b) Close-ups of the graphene/electrolyte interface for $\rho=$ 0.12, 0.31, 0.68, and 0.89 electrolyte systems are shown. (c) Illustration of the coarse-grained models used in the simulations for ACN (blue), [BMIM$^+$] (red), and [PF6$^-$] (green). The electrode is composed of three layers of carbon atoms with its basal plane exposed.}
\label{Fi:1}
\end{center} 
\end{figure*}

Using MD simulations in the constant potential ensemble, we can investigate the relationship between the molecular structure of the electrolyte and the thermal electrode charge fluctuations. This approach offers a physically transparent way to decompose the effects of microscopic correlations on electrochemical response\cite{limmer2013charge}. The subtle effects resulting from the interplay between these solvent-solvent, ion-ion and solvent-ion correlations are not easily understood from continuum treatments. The specific system considered here utilizes coarse-grained molecular models with nonspecific surface-fluid interactions and is aimed at capturing the behavior of a typical low molecular weight ionic liquid-solvent mixture in contact with idealized electrodes. Specifically, we employ molecular simulation models of butlymethylimidazolium hexafluorophosphate ([BMIM$^+$][PF6$^-$]) \cite{roy2010improved} - acetonitrile (ACN) \cite{edwards1984computer} mixtures bounded by electrodes modelled as three parallel ideal conductor honeycomb lattices of carbon atoms  on both sides \cite{cole1983interaction} as depicted in Fig. \ref{Fi:1}(a). Despite its relative simplicity, this model has been shown to yield good agreement between simulation and experiment for a variety of bulk and interfacial properties \cite{merlet2014electric,merlet2012new,pean2014dynamics,maroncelli2012measurements}. Figure \ref{Fi:1}(b) shows characteristic snapshots of the electrode-electrolyte interface for different ion-solvent compositions. Details on the molecular models are given in the Supporting Information.

The algorithm we use to maintain a constant potential across the capacitor follows from Reed et al. \cite{reed2007electrochemical} based on the work of Siepmann and Sprik \cite{siepmann1995influence}. During the simulation, the charge on each electrode atom fluctuates in response to the thermal motion of the electrolyte with fixed potential difference  $\Delta \Psi$, temperature $T$ and system volume $V$. The number of electrolyte molecules $N=N_\mathrm{i}+N_\mathrm{s}$, where $N_\mathrm{i}$ is the number of ions and $N_\mathrm{s}$ is the number of solvent molecules, is also kept fixed during the simulation. The electrode charges are determined at each time step by minimizing the potential energy subject to a constraint of constant voltage, which can be solved efficiently by matrix inversion\cite{gingrich2010ewald}. 
%This condition means that the electrodes polarize like a metal, where the potential inside the electrode is constant.
Within this ensemble, the differential capacitance, $C(\Delta \Psi)$, is calculated from the variance of electrode charge fluctuations using the fluctuation-dissipation theorem \cite{johnson1928thermal,nyquist1928thermal},
\begin{equation}
C(\Delta \Psi) =  \frac{\partial Q}{\partial \Delta \Psi} = \beta \langle \delta Q^2 \rangle \, ,
\end{equation}
where $Q$ is the total charge of one electrode, $\beta = 1/k_\mathrm{B} T$, with $k_\mathrm{B}$ Boltzmann's constant,  $\langle \dots \rangle$ indicates the ensemble average with constant $N$, $\Delta \Psi$ and $T=400$ K, and $\delta Q = Q-\langle Q\rangle$.

\begin{figure}[t]
\begin{center}
\includegraphics[width=8.5cm]{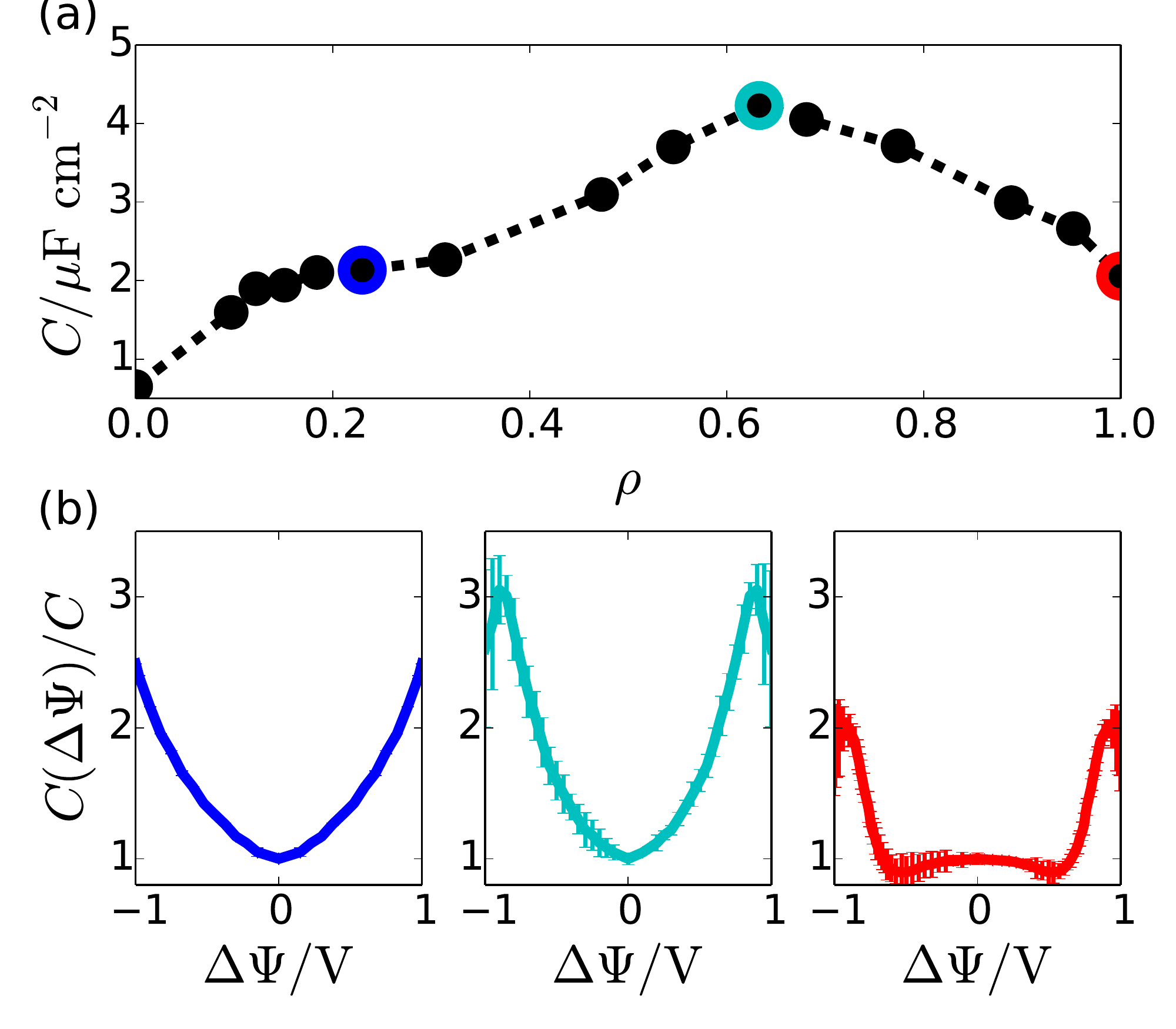}
\caption{Capacitance as a function of ion fraction and potential. (a) Capacitance at zero applied potential as a function of ion concentration, normalized by the area of the electrode. The line is a guide to the eye. Error estimates are smaller than the circle size. (b) Capacitance as a function of applied potential for three ionic liquid mole fractions, 0.23, 0.63, and 1 (left to right).}
\label{Fi:2}
\end{center} 
\end{figure}

Figure \ref{Fi:2}(a) shows $C$, the capacitance at zero applied potential, as a function of ion fraction $\rho = N_\mathrm{i}/N$. While $C$ increases with increasing ion concentration near $\rho = 0$, in the concentrated regime it decreases with ion concentration and exhibits a peak near $\rho = 0.63$. The increase in capacitance with increasing ion concentration in the dilute regime is expected from Gouy-Chapman-Stern theory, where near the potential of zero charge, the capacitance is proportional to the square root of ion concentration\cite{parsons1990electrical}. The peak in capacitance is consistent with experimental results of a different ionic liquid in contact with a molecularly rough electrode\cite{bozym2015anomalous}, suggesting an origin for this behavior within the electrolyte.

Figure \ref{Fi:2}(b) shows the capacitance as a function of potential calculated using histogram reweighting techniques\cite{limmer2013charge}. Details on the methods are in the Supporting Information. Capacitance profiles as a function of electrode potential for the three systems in Fig. \ref{Fi:2}(b) exhibit a broadening near the potential of zero charge with increasing concentration, consistent with experiment\cite{bozym2015anomalous}. The nonmonotonic concentration dependence of capacitance at $\Delta \Psi = 0$ V is observed throughout the $2$ V potential window. For pure ionic liquids, previous studies foreshadowed an unbounded capacitance at $\Delta \Psi = \pm 0.9$ V due to a surface phase transition\cite{merlet2014electric}. In that regime, finite size effects not studied here are likely important. The capacitance calculated from electrode charge fluctuations should be symmetric around $\Delta \Psi = 0$ V, and any deviation is due to statistical uncertainty.

The capacitance-voltage relationship of pure [BMIM$^+$][PF6$^-$] features symmetric double-peaks \cite{fedorov2014ionic}. When the ionic solution is diluted to $\rho = 0.63$, the peaks are retained and curvature around $\Delta \Psi = 0$ V also increases. The peaks at moderate to high voltages result from the small bias to expel counterions and solvent from the interface\cite{bazant2011double,fedorov2014ionic,kornyshev2007double} with increasing potential difference. At larger potentials when the ionic adlayer condenses, charge fluctuations are sterically suppressed and capacitance decreases\cite{bazant2011double,fedorov2014ionic,kornyshev2007double}. As $\rho$ is decreased further, the curve becomes U-shaped. This U-shape at low ion concentrations results from charge fluctuations that are proportional to the number of ions near the interface, whose number can grow by expelling solvent molecules away from the electrode without the steric constraints that occur at high concentrations\cite{bazant2011double,fedorov2014ionic,kornyshev2007double}. Hereafter, we focus on electronic and structural properties at $\Delta \Psi = 0$ V.

The concentration dependence of the capacitance can be understood by decomposing the charge fluctuations into different components,
\begin{equation}
\label{Eq:fluct}
\langle (\delta Q)^2 \rangle = \langle (\delta Q_\mathrm{s}^2) \rangle + \langle (\delta Q_\mathrm{i})^2 \rangle+2 \langle \delta Q_\mathrm{s}\delta Q_\mathrm{i} \rangle \, ,
\end{equation}
where $\langle (\delta Q_\mathrm{s})^2  \rangle$ and $\langle (\delta Q_\mathrm{i})^2  \rangle$ are due the the solvent and ions respectively,
and $\langle \delta Q_\mathrm{s}\delta Q_\mathrm{i} \rangle$ are fluctuations induced by ion-solvent correlations. This decomposition is possible for a conductor where the electrode charges are linear functions of the partial charges of the electrolyte\cite{limmer2013charge}. Microscopically, they can be explicitly computed using the Stillinger-Lovett sum rules for the charge-charge structure factor\cite{stillinger1968general}. Simulations of a pure ACN-electrode system gives $\beta \langle (\delta Q)^2 \rangle = \beta \langle (\delta Q_\mathrm{s})^2 \rangle = 0.65 \pm 0.06$ $\mu$F/cm$^2$, which is significantly smaller than that of the ionic solutions. Therefore, $\langle (\delta Q_\mathrm{s})^2 \rangle$ are expected to be negligible over the entire concentration regime, consistent with the expectation that since solvent molecules do not carry a net charge they cannot efficiently polarize the electrode surface.

The second term in Eq. \ref{Eq:fluct},  $\langle (\delta Q_\mathrm{i})^2$, is expected to govern the behavior of electrode charge fluctuations both in the dilute ion regime and as $\rho \rightarrow 1$. For an ideal solution, the differential capacitance can only increase with an increase in the fraction of independent ions, a number set by the screening length. Deviations from this monotonic increase are expected in the limit of a pure ionic solution, as ions cease behaving ideally. However as Fig. \ref{Fi:2}(a) shows, rather than a plateau in the capacitance as $\rho \rightarrow 1$, there is a maximum at intermediate concentrations. This suggests that the third term, $\langle \delta Q_\mathrm{s}\delta Q_\mathrm{i} \rangle$, plays a significant role in determining the magnitude of charge fluctuations. In fact, as discussed below, for concentrated electrolytes the motion of solvent molecules is highly correlated with the charge density fluctuations near the electrode interface, due to the incompressibility of the solution. These ion-solvent correlations enhance the electrode charge fluctuations by affecting the magnitude of the induced image charge on the electrodes.

\begin{figure}[b]
\begin{center}
\includegraphics[width=8.5cm]{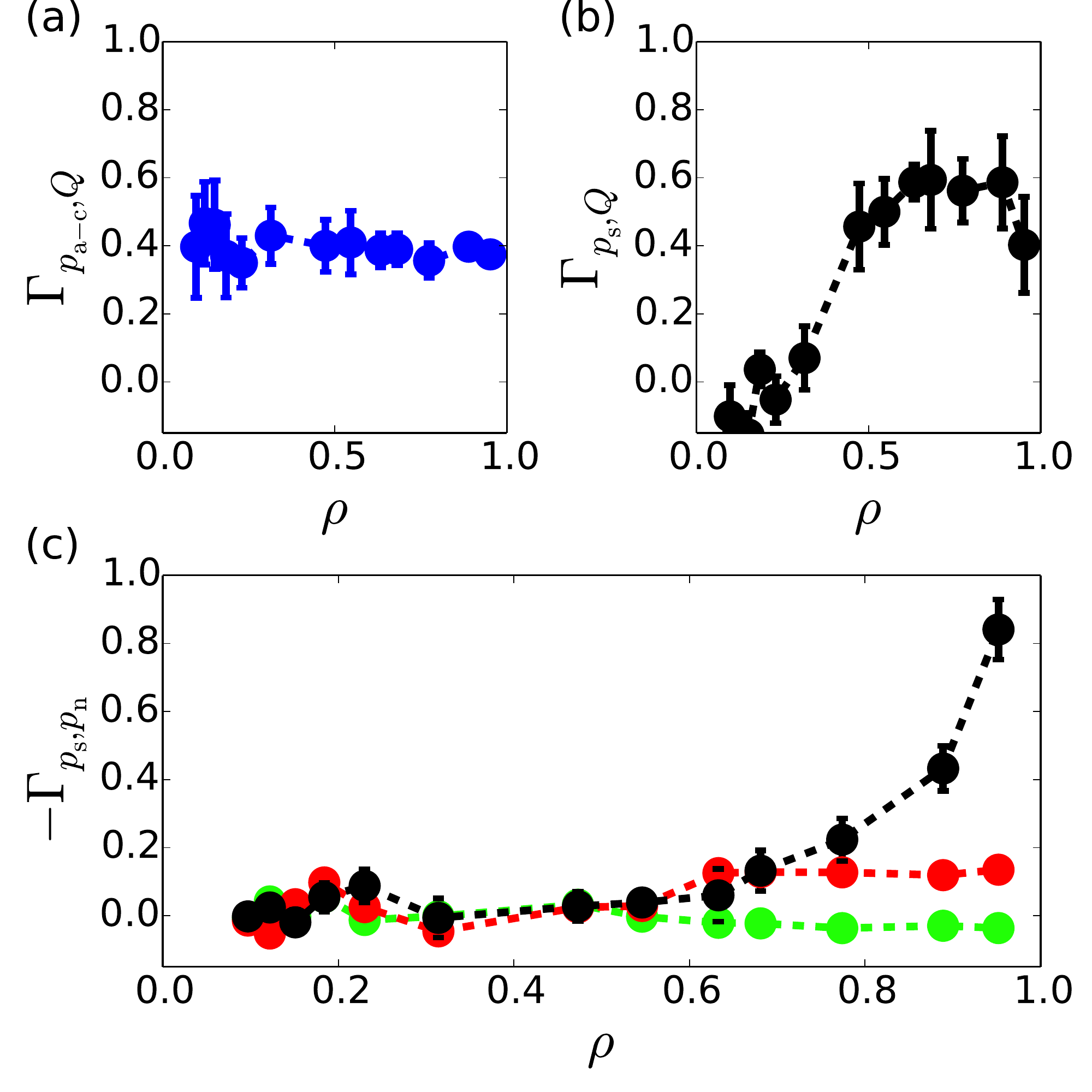}
\caption{Composition dependence of electrolyte-electrode charge correlations at $\Delta \Psi = 0$ V. Normalized static cross-correlation coefficients between (a) electrode charge fluctuations and ion polarization; (b) electrode charge fluctuations and interfacial solvent concentration weighted by its displacement from the electrode; (c) interfacial solvent concentration weighted by its displacement from the electrode and ion polarization $p_\mathrm{a-c}$ (black), cation $p_\mathrm{c}$ (red) or anion $p_\mathrm{a}$ (green) concentration, weighted by its displacement from the electrode at the interface. The lines are guides to the eye. Error bars represent one standard deviation and are smaller than the markers when not shown.}
\label{Fi:3}
\end{center} 
\end{figure}

In particular, Fig. \ref{Fi:3}(a) illustrates the concentration dependence of the cross correlation coefficient between electrode charge fluctuations and ion polarization,
\begin{equation}
\Gamma_{p_{x},Q} = \frac{\langle \delta p_{x} \delta Q\rangle}{\sqrt{\langle (\delta p_{x})^2\rangle \langle (\delta Q)^2\rangle}} \, ,
\end{equation}
where $p_x = \{p_\mathrm{a-c},p_\mathrm{s}\}$, and $p_\mathrm{a-c}$ is the charge polarization near the electrode and $p_\mathrm{s}$ is the interfacial solvent concentration, weighted by its displacement from the electrode. The interfacial polarization is computed by summing over the $N_\mathrm{i}$ ions with instantaneous displacement from the electrode less than $z_\mathrm{c} = 0.9$ nm, as
\begin{equation}
\label{Eq:rhos}
p_\mathrm{a-c} = \frac{1}{v} \sum_{i=1}^{N_\mathrm{i}} \hat{z}_i q_i \Theta(z_\mathrm{c} - \hat{z}_i) \, ,
\end{equation}
and similarly, 
\begin{equation}
\label{Eq:rhoss}
p_\mathrm{s} = \frac{1}{v} \sum_{i=1}^{N_\mathrm{s}} \hat{z}_i  \Theta(z_\mathrm{c} - \hat{z}_i) \, ,
\end{equation}
where $q_i$ is the charge of ion $i$, $\Theta[x]$ is the Heaviside step function, $\hat{z}_i$ is the instantaneous displacement of the ion from the electrode and $v$ is the volume of the $L\times L \times z_\mathrm{c}$ slab. The thickness $z_\mathrm{c}$ is chosen to accommodate the first two solvation layers near the electrode and the results are qualitatively insensitive to its precise value. The constant profile in Fig. \ref{Fi:3}(a) reveals that ion polarization at the interface has a similar effect on electrode polarization regardless of the electrolyte composition. Correlations between interfacial solvent displacement and electrode charge become more pronounced with increasing ion concentration. This is indicated in Fig. \ref{Fi:3}(b) by the cross correlation coefficient between $Q$ and $p_\mathrm{s}$.

Since solvent fluctuations themselves cannot significantly polarize the electrode, increasing $\Gamma_{p_\mathrm{s},Q}$ with concentration implies that solvent motion is correlated with ion polarization that in turn gives rise to the increase in Fig. \ref{Fi:3}(b). In fact, Fig. \ref{Fi:3}(c) shows solvent fluctuations correlated with cation and anion center of mass and ion polarization fluctuations,
\begin{equation}
\Gamma_{p_\mathrm{s},p_\mathrm{n}} = \frac{\langle \delta p_\mathrm{s} \delta p_\mathrm{n}\rangle}{\sqrt{\langle (\delta p_\mathrm{s})^2\rangle \langle (\delta p_\mathrm{n})^2\rangle}}
\end{equation}
where $p_\mathrm{n} = \{p_\mathrm{a},p_\mathrm{c},p_\mathrm{a-c}\}$ and $p_\mathrm{a}$ $(p_\mathrm{c})$ is the interfacial anion (cation) concentration, weighted by its displacement from the electrode, and calculated analogously to $p_\mathrm{s}$. While cation-solvent and anion-solvent correlations do not exhibit strong compositional dependence, solvent-polarization correlations are enhanced with increasing ion concentration. The negative value of this covariance arises molecularly from swapping motions that simultaneously moves the center of mass of the solvent molecules away from the electrode while polarizing the electrode by increasing the charge separation in the direction of the electrode. 

The source of these increasing ion-solvent correlations can be understood as arising from the solvent's ability to facilitate fluctuations in an otherwise dense, incompressible and strongly associated fluid. Namely, for an ion pair to be separated and polarize the electrode, a fluctuation in the surrounding solvent must occur to stabilize that polarization. The increase in magnitude of this correlation with ion concentration, results from the increasing steric constraints of the ions. This picture is consistent with the results from lattice model calculations\cite{bozym2015anomalous}, where the capacitance maximum can be recovered by treating the solvent molecule as a defect that enables ionic reorganization, but does not directly contribute to the charge fluctuations.

\begin{figure}[b]
\begin{center}
\includegraphics[width=8.5cm]{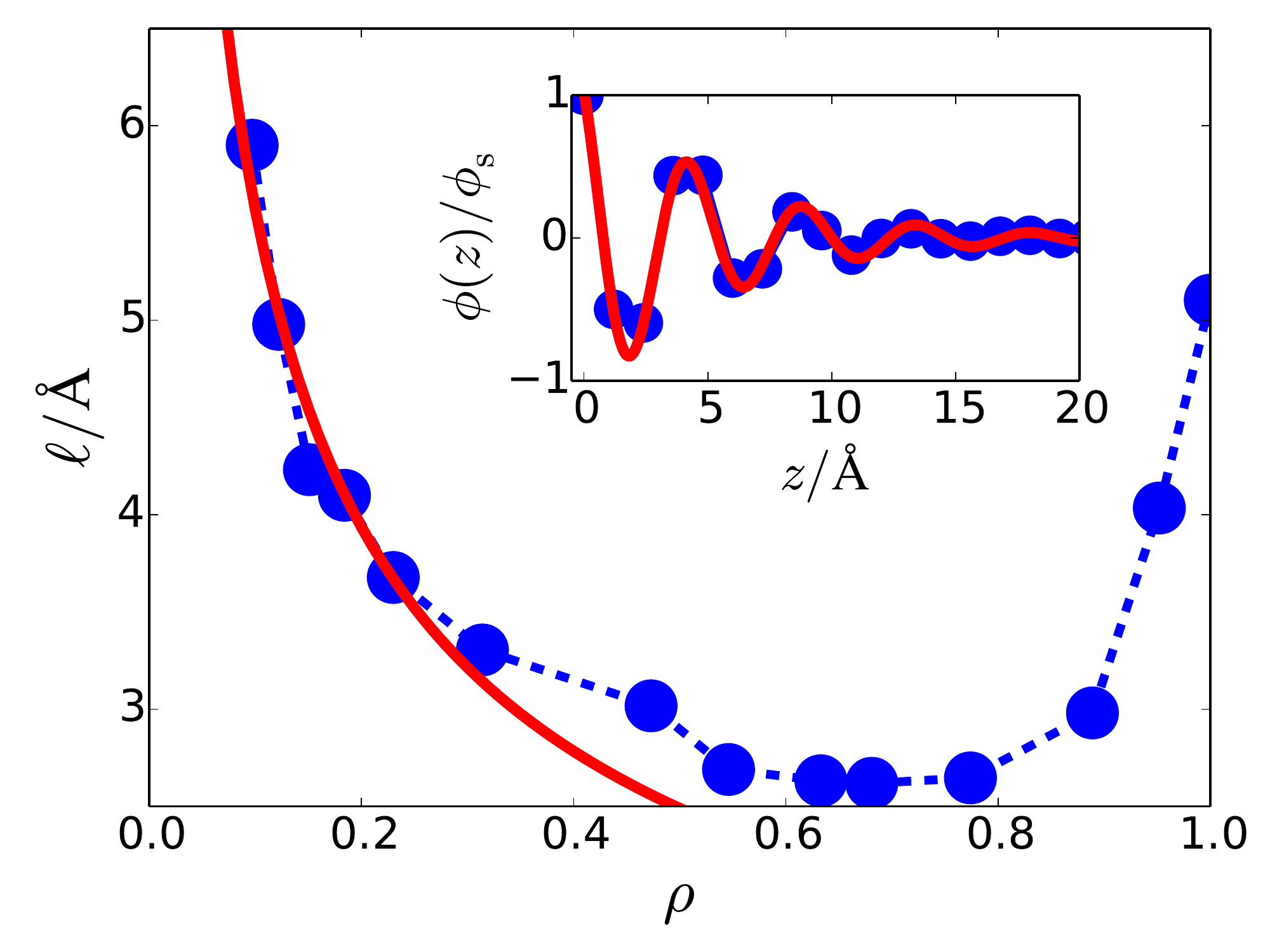}
\caption{Charge density correlation length as a function of ion concentration. See Eq. \ref{Eq:Phi} for definition. The solid line is a fit to $\ell \propto 1/ \sqrt{\rho}$, as predicted from Gouy-Chapman theory. The dashed line is a guide to the eye. Coarse-grained charge density as a function of distance from the electrode surface is given in the inset. The solid line is a fit to Eq. \ref{Eq:Phi}.}
\label{Fi:4}
\end{center} 
\end{figure}

In order to provide a structural interpretation of the composition dependence of the correlations described above, we construct a coarse-grained charge density distribution away from the electrode. Specifically, we take the out-of-plane charge density distribution computed from the simulations and average over 1 $\mathrm{\AA}$ bins, so as to integrate out small length scale features associated with the internal structure of the ions. A representative coarse-grained profile is plotted in the inset of Fig. \ref{Fi:4}. For all concentrations studied, the coarse-grained charge density can be fit with a damped harmonic function with decay constant $\ell$ and periodicity $q_\mathrm{s}$,
\begin{equation}
\label{Eq:Phi}
\phi(z) = \phi_\mathrm{s} e^{-z/\ell} \cos (2 \pi z/q_\mathrm{s} + \theta)
\end{equation}
where $z$ is distance from the first maximum of $\phi(z)$, $\phi_\mathrm{s}$ is the magnitude of that first maximum and $\theta$ is an angular offset. This functional form has been derived theoretically for pure ionic liquids \cite{limmer2015interfacial} and is routinely used in experimental studies to fit the charge density profile of dense electrolytes\cite{mezger2015solid}. The periodicity of charge oscillations, $q_\mathrm{s}$ originates from the interplay of excluded volume of the ions and screening\cite{limmer2015interfacial} and we find that it can be fixed to 4.2 $\mathrm{\AA}$, or a little smaller than the average size of the ions, for all concentrations. This indicates that ions can maintain their preferred distance from each other regardless of electrolyte composition.

The decay length, $\ell$, reflects the scale of ionic correlations away from the electrode surface. As shown in Fig. \ref{Fi:4}, in the dilute regime $1/\ell$ scales as the square root of ion concentration, qualitatively in agreement with Debye-H\"uckel theory, though quantitatively inconsistent with the known dielectric constant for this solvent model \cite{edwards1984computer}. In the pure ionic liquid, $\ell$ is determined by steric interactions \cite{limmer2015interfacial} and thus, its decrease  with decreasing ion concentration signifies a solvent mediated reduction in packing constraints. This behavior is consistent with an $\ell \propto \sqrt{1-\rho}$ dependence extracted from lattice model calculations\cite{bozym2015anomalous} where the capacitance enhancement in the concentrated regime is facilitated by the solvent's ability to enable ionic reorganization in an otherwise incompressible fluid. This reduction in charge density oscillations with solvent has been noted in previous simulations of different solutions \cite{merlet2013influence}. The steeper charge density decay as a function of distance from the electrode surface with the dilution of the ionic solution in the concentrated regime is analogous to that found in experiment \cite{mezger2015solid}. Both $\ell$ and $q_\mathrm{s}$ obtained from the interfacial profiles show similar trends with the correlation length and periodicity extracted from bulk radial charge density distributions as given in Supporting Information, consistent with the expectation that the electrode interacts with the solutions weakly.

\begin{figure}
\begin{center}
\includegraphics[width=8.5cm]{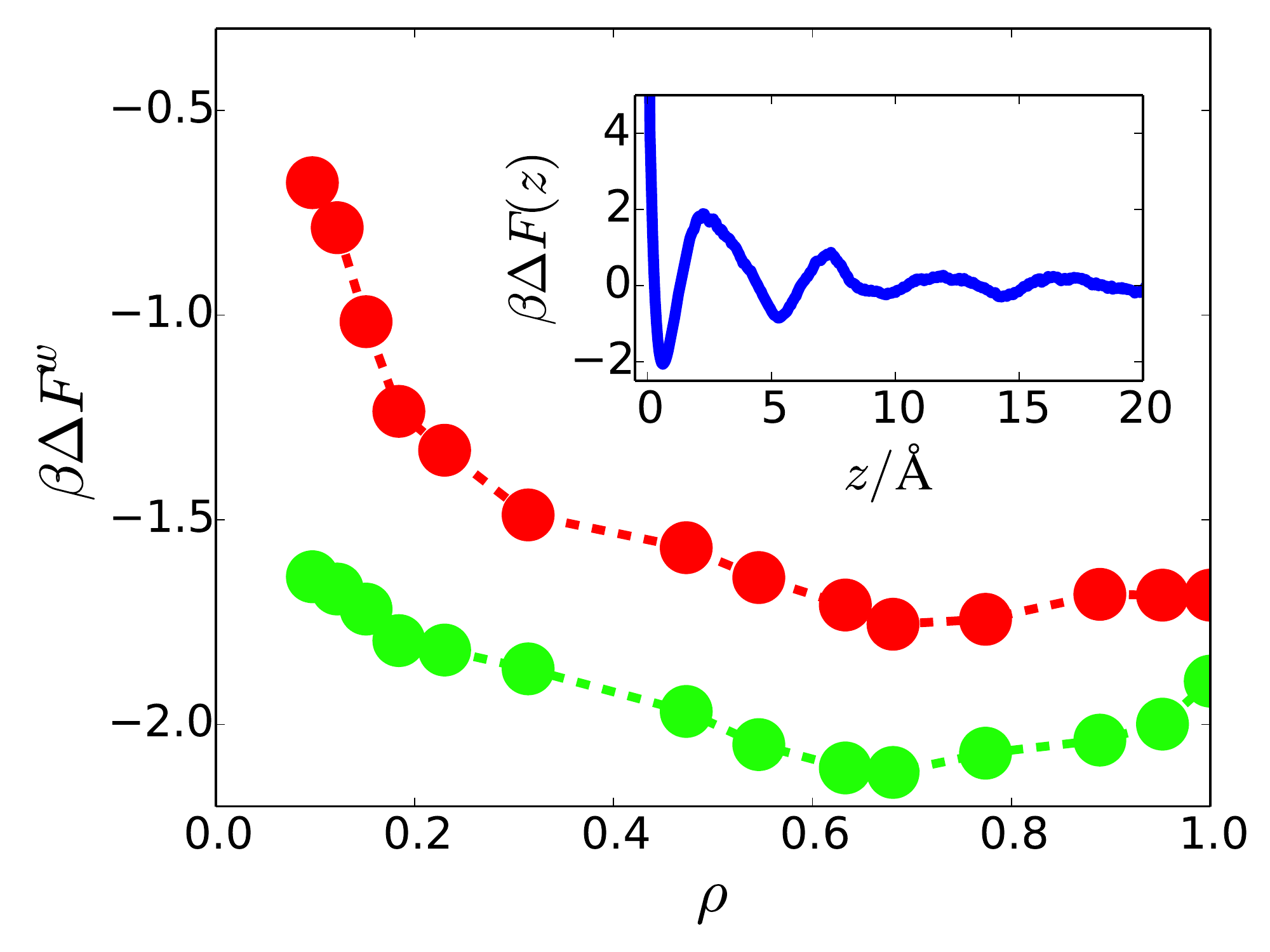}
\caption{Ion surface adsorption free energies at $\Delta \Psi = 0$ V. The adsorption free energy obtained for the cation (red), and anion (green), as a function of composition. Error estimates are smaller than the circle size. The lines are guides to the eye. The inset shows the free energy profile for moving an anion at $\rho = 0.09$ as a function of distance from the electrode surface.}
\label{Fi:5}
\end{center} 
\end{figure}

The maximum charge density near the electrode, $\phi_\mathrm{s}$, is determined by the relative surface propensities of cations, anions and solvent and depends intimately on the details of the intermolecular interactions \cite{dos2008consistency,otten2012elucidating,liu2014strong,jungwirth2006specific}. To quantify this, the inset of Fig. \ref{Fi:5} depicts the free energy difference, $\Delta F(z)$, for moving an ion from the bulk to a distance $z$ from the electrode, computed from
\begin{equation}
\beta \Delta F(z) = - \ln \left [ \rho(z) / \rho_\mathrm{b} \right ]
\end{equation}
where $\rho(z)$ is the local ion density and $\rho_\mathrm{b}$ is the bulk ion density. Figure \ref{Fi:5} shows the depths of the first minimum, $\Delta F^w$, for both cation and anion, which is indicative of the strength of selective ion adsorption at the interface.

Adsorption of both cations and anions depends strongly on the bulk electrolyte concentration. In the dilute regime, the affinity of both ion types for the interface increases with ion fraction. The effect is larger for the cation, consistent with previous work suggesting that it is weakly solvated \cite{merlet2013influence}. In the concentrated regime, the adsorption affinities of both ions do not change appreciably, though the anion is slightly depleted from the interface in the pure ionic liquid. These observations mirror those derived from the charge density distribution. In the dilute regime, changing ion concentration changes the average density of ions near the interface, which acts to increase fluctuations proportionally. In the concentrated regime, average densities are not strongly affected by solvent concentration, but fluctuations around the mean are influenced. In the charge density, this is manifested in the extent of charge density layering, while here it is manifested in the changing barrier height to move between layers. These observations are in accord with the results of Feng et al.\cite{Feng2014Water} where the introduction of a small amount of water into an ionic liquid system was found to have very little impact on ion adsorption affinities but leads to an increase in the capacitance at 2V. The authors show that the capacitance enhancement is associated with the larger potential drop at the electrical double layer in the absence of water molecules, which disrupts the ion ordering at the interface, and stems mainly from energetic effects. Our studies in progress reveal that in the acetonitrile-ionic liquid system entropic effects also play an important role due to the fact that the solvent molecules are comparable in size to the ions.

Molecular dynamics simulation in a constant potential ensemble allows the differential capacitance to be computed and interpreted as a fluctuation quantity. This has enabled an intuitive decomposition of the capacitance, and associated maximum at intermediate concentrations, in terms of different molecular contributions. These results provide insights into the microscopic mechanisms that determine the electronic properties of the interface and enable means to examine how molecular features such as polarity, size, and shape affect electrochemical responses.
\subsection*{Acknowledgements}
We would like to thank Benjamin Rotenberg for useful comments on an earlier version of this manuscript. This work was supported in part by a Multidisciplinary University Research Initiative (MURI) through the Air Force Office of Scientific Research (Grant FA9550-13-1-0004). D.T.L. was supported by the Princeton Center for Theoretical Science. P.G.D. acknowledges the support of the National Science Foundation (Grant Nos. CBET-1263565 and CHE-1213343).
\subsection*{Supporting Information}
Details on the molecular simulation models, histogram reweighting techniques used to compute capacitance as a continuous function of potential, and bulk radial charge density distributions
\subsection*{References }

\bibliographystyle{biochem}  
\bibliography{ref}

\providecommand{\latin}[1]{#1}
\providecommand*\mcitethebibliography{\thebibliography}
\csname @ifundefined\endcsname{endmcitethebibliography}
  {\let\endmcitethebibliography\endthebibliography}{}
\begin{mcitethebibliography}{44}
\providecommand*\natexlab[1]{#1}
\providecommand*\mciteSetBstSublistMode[1]{}
\providecommand*\mciteSetBstMaxWidthForm[2]{}
\providecommand*\mciteBstWouldAddEndPuncttrue
  {\def\EndOfBibitem{\unskip.}}
\providecommand*\mciteBstWouldAddEndPunctfalse
  {\let\EndOfBibitem\relax}
\providecommand*\mciteSetBstMidEndSepPunct[3]{}
\providecommand*\mciteSetBstSublistLabelBeginEnd[3]{}
\providecommand*\EndOfBibitem{}
\mciteSetBstSublistMode{f}
\mciteSetBstMaxWidthForm{subitem}{(\alph{mcitesubitemcount})}
\mciteSetBstSublistLabelBeginEnd
  {\mcitemaxwidthsubitemform\space}
  {\relax}
  {\relax}

\bibitem[Bozym \latin{et~al.}(2015)Bozym, Uralcan, Limmer, Pope, Szamreta,
  Debenedetti, and Aksay]{bozym2015anomalous}
Bozym,~D.~J., Uralcan,~B., Limmer,~D.~T., Pope,~M.~A., Szamreta,~N.~J.,
  Debenedetti,~P.~G., and Aksay,~I.~A. (2015) Anomalous Capacitance Maximum of
  the Glassy Carbon--Ionic Liquid Interface through Dilution with Organic
  Solvents. \emph{The journal of physical chemistry letters} \emph{6},
  2644--2648\relax
\mciteBstWouldAddEndPuncttrue
\mciteSetBstMidEndSepPunct{\mcitedefaultmidpunct}
{\mcitedefaultendpunct}{\mcitedefaultseppunct}\relax
\EndOfBibitem
\bibitem[Liu \latin{et~al.}(2012)Liu, Yan, Lang, and Xue]{liu2012effects}
Liu,~W., Yan,~X., Lang,~J., and Xue,~Q. (2012) Effects of Concentration and
  Temperature of EMIMBF 4/Acetonitrile Electrolyte on the Supercapacitive
  Behavior of Graphene Nanosheets. \emph{Journal of Materials Chemistry}
  \emph{22}, 8853--8861\relax
\mciteBstWouldAddEndPuncttrue
\mciteSetBstMidEndSepPunct{\mcitedefaultmidpunct}
{\mcitedefaultendpunct}{\mcitedefaultseppunct}\relax
\EndOfBibitem
\bibitem[Siinor \latin{et~al.}(2012)Siinor, Siimenson, Ivani{\v{s}}t{\v{s}}ev,
  Lust, and Lust]{siinor2012influence}
Siinor,~L., Siimenson,~C., Ivani{\v{s}}t{\v{s}}ev,~V., Lust,~K., and Lust,~E.
  (2012) Influence of Cation Chemical Composition and Structure on the Double
  Layer Capacitance for Bi (111)| Room Temperature Ionic Liquid Interface.
  \emph{Journal of Electroanalytical Chemistry} \emph{668}, 30--36\relax
\mciteBstWouldAddEndPuncttrue
\mciteSetBstMidEndSepPunct{\mcitedefaultmidpunct}
{\mcitedefaultendpunct}{\mcitedefaultseppunct}\relax
\EndOfBibitem
\bibitem[Fedorov and Kornyshev(2014)Fedorov, and Kornyshev]{fedorov2014ionic}
Fedorov,~M.~V., and Kornyshev,~A.~A. (2014) Ionic Liquids at Electrified
  Interfaces. \emph{Chemical reviews} \emph{114}, 2978--3036\relax
\mciteBstWouldAddEndPuncttrue
\mciteSetBstMidEndSepPunct{\mcitedefaultmidpunct}
{\mcitedefaultendpunct}{\mcitedefaultseppunct}\relax
\EndOfBibitem
\bibitem[Kornyshev(2007)]{kornyshev2007double}
Kornyshev,~A.~A. (2007) Double-Layer in Ionic Liquids: Paradigm Change?
  \emph{The Journal of Physical Chemistry B} \emph{111}, 5545--5557\relax
\mciteBstWouldAddEndPuncttrue
\mciteSetBstMidEndSepPunct{\mcitedefaultmidpunct}
{\mcitedefaultendpunct}{\mcitedefaultseppunct}\relax
\EndOfBibitem
\bibitem[Fedorov and Kornyshev(2008)Fedorov, and Kornyshev]{fedorov2008towards}
Fedorov,~M.~V., and Kornyshev,~A.~A. (2008) Towards Understanding the Structure
  and Capacitance of Electrical Double Layer in Ionic Liquids.
  \emph{Electrochimica Acta} \emph{53}, 6835 -- 6840\relax
\mciteBstWouldAddEndPuncttrue
\mciteSetBstMidEndSepPunct{\mcitedefaultmidpunct}
{\mcitedefaultendpunct}{\mcitedefaultseppunct}\relax
\EndOfBibitem
\bibitem[Fedorov and Kornyshev(2008)Fedorov, and Kornyshev]{fedorov2008ionic}
Fedorov,~M.~V., and Kornyshev,~A.~A. (2008) Ionic Liquid Near a Charged Wall:
  Structure and Capacitance of Electrical Double Layer. \emph{The Journal of
  Physical Chemistry B} \emph{112}, 11868--11872\relax
\mciteBstWouldAddEndPuncttrue
\mciteSetBstMidEndSepPunct{\mcitedefaultmidpunct}
{\mcitedefaultendpunct}{\mcitedefaultseppunct}\relax
\EndOfBibitem
\bibitem[Fedorov and Kornyshev(2009)Fedorov, and Kornyshev]{fedorov2008err}
Fedorov,~M.~V., and Kornyshev,~A.~A. (2009) Ionic Liquid Near a Charged Wall:
  Structure and Capacitance of Electrical Double Layer. \emph{The Journal of
  Physical Chemistry B} \emph{113}, 4500--4500\relax
\mciteBstWouldAddEndPuncttrue
\mciteSetBstMidEndSepPunct{\mcitedefaultmidpunct}
{\mcitedefaultendpunct}{\mcitedefaultseppunct}\relax
\EndOfBibitem
\bibitem[Bazant \latin{et~al.}(2011)Bazant, Storey, and
  Kornyshev]{bazant2011double}
Bazant,~M.~Z., Storey,~B.~D., and Kornyshev,~A.~A. (2011) Double Layer in Ionic
  Liquids: Overscreening versus Crowding. \emph{Physical Review Letters}
  \emph{106}, 046102\relax
\mciteBstWouldAddEndPuncttrue
\mciteSetBstMidEndSepPunct{\mcitedefaultmidpunct}
{\mcitedefaultendpunct}{\mcitedefaultseppunct}\relax
\EndOfBibitem
\bibitem[Bazant \latin{et~al.}(2012)Bazant, Storey, and
  Kornyshev]{bazant2011err}
Bazant,~M.~Z., Storey,~B.~D., and Kornyshev,~A.~A. (2012) Erratum: Double Layer
  in Ionic Liquids: Overscreening versus Crowding [Phys. Rev. Lett.
  \textbf{106} , 046102 (2011)]. \emph{Phys. Rev. Lett.} \emph{109},
  149903\relax
\mciteBstWouldAddEndPuncttrue
\mciteSetBstMidEndSepPunct{\mcitedefaultmidpunct}
{\mcitedefaultendpunct}{\mcitedefaultseppunct}\relax
\EndOfBibitem
\bibitem[Wu \latin{et~al.}(2011)Wu, Jiang, Jiang, Jin, and
  Henderson]{wu2011classical}
Wu,~J., Jiang,~T., Jiang,~D., Jin,~Z., and Henderson,~D. (2011) A Classical
  Density Functional Theory for Interfacial Layering of Ionic Liquids.
  \emph{Soft Matter} \emph{7}, 11222--11231\relax
\mciteBstWouldAddEndPuncttrue
\mciteSetBstMidEndSepPunct{\mcitedefaultmidpunct}
{\mcitedefaultendpunct}{\mcitedefaultseppunct}\relax
\EndOfBibitem
\bibitem[Simon and Gogotsi(2008)Simon, and Gogotsi]{simon2008materials}
Simon,~P., and Gogotsi,~Y. (2008) Materials for Electrochemical Capacitors.
  \emph{Nature materials} \emph{7}, 845--854\relax
\mciteBstWouldAddEndPuncttrue
\mciteSetBstMidEndSepPunct{\mcitedefaultmidpunct}
{\mcitedefaultendpunct}{\mcitedefaultseppunct}\relax
\EndOfBibitem
\bibitem[Miller and Simon(2008)Miller, and Simon]{miller2008electrochemical}
Miller,~J.~R., and Simon,~P. (2008) Electrochemical Capacitors for Energy
  Management. \emph{Science Magazine} \emph{321}, 651--652\relax
\mciteBstWouldAddEndPuncttrue
\mciteSetBstMidEndSepPunct{\mcitedefaultmidpunct}
{\mcitedefaultendpunct}{\mcitedefaultseppunct}\relax
\EndOfBibitem
\bibitem[Mezger \latin{et~al.}(2015)Mezger, Roth, Schr{\"o}der, Reichert,
  Pontoni, and Reichert]{mezger2015solid}
Mezger,~M., Roth,~R., Schr{\"o}der,~H., Reichert,~P., Pontoni,~D., and
  Reichert,~H. (2015) Solid-Liquid Interfaces of Ionic Liquid
  Solutions—Interfacial Layering and Bulk Correlations. \emph{The Journal of
  chemical physics} \emph{142}, 164707\relax
\mciteBstWouldAddEndPuncttrue
\mciteSetBstMidEndSepPunct{\mcitedefaultmidpunct}
{\mcitedefaultendpunct}{\mcitedefaultseppunct}\relax
\EndOfBibitem
\bibitem[Mezger \latin{et~al.}(2009)Mezger, Schramm, Schr{\"o}der, Reichert,
  Deutsch, De~Souza, Okasinski, Ocko, Honkim{\"a}ki, and
  Dosch]{mezger2009layering}
Mezger,~M., Schramm,~S., Schr{\"o}der,~H., Reichert,~H., Deutsch,~M.,
  De~Souza,~E.~J., Okasinski,~J.~S., Ocko,~B.~M., Honkim{\"a}ki,~V., and
  Dosch,~H. (2009) Layering of [BMIM]+-based Ionic Liquids at a Charged
  Sapphire Interface. \emph{The Journal of chemical physics} \emph{131},
  094701\relax
\mciteBstWouldAddEndPuncttrue
\mciteSetBstMidEndSepPunct{\mcitedefaultmidpunct}
{\mcitedefaultendpunct}{\mcitedefaultseppunct}\relax
\EndOfBibitem
\bibitem[Hayes \latin{et~al.}(2011)Hayes, Borisenko, Tam, Howlett, Endres, and
  Atkin]{hayes2011double}
Hayes,~R., Borisenko,~N., Tam,~M.~K., Howlett,~P.~C., Endres,~F., and Atkin,~R.
  (2011) Double Layer Structure of Ionic Liquids at the Au (111) Electrode
  Interface: An Atomic Force Microscopy Investigation. \emph{The Journal of
  Physical Chemistry C} \emph{115}, 6855--6863\relax
\mciteBstWouldAddEndPuncttrue
\mciteSetBstMidEndSepPunct{\mcitedefaultmidpunct}
{\mcitedefaultendpunct}{\mcitedefaultseppunct}\relax
\EndOfBibitem
\bibitem[Atkin \latin{et~al.}(2011)Atkin, Borisenko, Dr{\"u}schler, El~Abedin,
  Endres, Hayes, Huber, and Roling]{atkin2011situ}
Atkin,~R., Borisenko,~N., Dr{\"u}schler,~M., El~Abedin,~S.~Z., Endres,~F.,
  Hayes,~R., Huber,~B., and Roling,~B. (2011) An in Situ STM/AFM and Impedance
  Spectroscopy Study of the Extremely Pure 1-butyl-1-methylpyrrolidinium tris
  (pentafluoroethyl) trifluorophosphate/Au (111) interface: Potential Dependent
  Solvation Layers and the Herringbone Reconstruction. \emph{Physical Chemistry
  Chemical Physics} \emph{13}, 6849--6857\relax
\mciteBstWouldAddEndPuncttrue
\mciteSetBstMidEndSepPunct{\mcitedefaultmidpunct}
{\mcitedefaultendpunct}{\mcitedefaultseppunct}\relax
\EndOfBibitem
\bibitem[Carstens \latin{et~al.}(2012)Carstens, Hayes, El~Abedin, Corr, Webber,
  Borisenko, Atkin, and Endres]{carstens2012situ}
Carstens,~T., Hayes,~R., El~Abedin,~S.~Z., Corr,~B., Webber,~G.~B.,
  Borisenko,~N., Atkin,~R., and Endres,~F. (2012) In Situ STM, AFM and DTS
  Study of the Interface 1-hexyl-3-methylimidazolium tris (pentafluoroethyl)
  trifluorophosphate/Au (111). \emph{Electrochimica Acta} \emph{82},
  48--59\relax
\mciteBstWouldAddEndPuncttrue
\mciteSetBstMidEndSepPunct{\mcitedefaultmidpunct}
{\mcitedefaultendpunct}{\mcitedefaultseppunct}\relax
\EndOfBibitem
\bibitem[Merlet \latin{et~al.}(2014)Merlet, Limmer, Salanne, Van~Roij, Madden,
  Chandler, and Rotenberg]{merlet2014electric}
Merlet,~C., Limmer,~D.~T., Salanne,~M., Van~Roij,~R., Madden,~P.~A.,
  Chandler,~D., and Rotenberg,~B. (2014) The Electric Double Layer Has a Life
  of Its Own. \emph{The Journal of Physical Chemistry C} \emph{118},
  18291--18298\relax
\mciteBstWouldAddEndPuncttrue
\mciteSetBstMidEndSepPunct{\mcitedefaultmidpunct}
{\mcitedefaultendpunct}{\mcitedefaultseppunct}\relax
\EndOfBibitem
\bibitem[Limmer(2015)]{limmer2015interfacial}
Limmer,~D.~T. (2015) Interfacial Ordering and Accompanying Divergent
  Capacitance at Ionic Liquid-Metal Interfaces. \emph{Physical review letters}
  \emph{115}, 256102\relax
\mciteBstWouldAddEndPuncttrue
\mciteSetBstMidEndSepPunct{\mcitedefaultmidpunct}
{\mcitedefaultendpunct}{\mcitedefaultseppunct}\relax
\EndOfBibitem
\bibitem[Arulepp \latin{et~al.}(2004)Arulepp, Permann, Leis, Perkson, Rumma,
  J{\"a}nes, and Lust]{arulepp2004influence}
Arulepp,~M., Permann,~L., Leis,~J., Perkson,~A., Rumma,~K., J{\"a}nes,~A., and
  Lust,~E. (2004) Influence of the Solvent Properties on the Characteristics of
  a Double Layer Capacitor. \emph{Journal of Power Sources} \emph{133},
  320--328\relax
\mciteBstWouldAddEndPuncttrue
\mciteSetBstMidEndSepPunct{\mcitedefaultmidpunct}
{\mcitedefaultendpunct}{\mcitedefaultseppunct}\relax
\EndOfBibitem
\bibitem[Pope and Aksay(2015)Pope, and Aksay]{Pope2015Four}
Pope,~M.~A., and Aksay,~I.~A. (2015) Four-Fold Increase in the Intrinsic
  Capacitance of Graphene through Functionalization and Lattice Disorder.
  \emph{The Journal of Physical Chemistry C} \emph{119}, 20369--20378\relax
\mciteBstWouldAddEndPuncttrue
\mciteSetBstMidEndSepPunct{\mcitedefaultmidpunct}
{\mcitedefaultendpunct}{\mcitedefaultseppunct}\relax
\EndOfBibitem
\bibitem[Feng \latin{et~al.}(2011)Feng, Huang, Sumpter, Meunier, and
  Qiao]{feng2011counter}
Feng,~G., Huang,~J., Sumpter,~B.~G., Meunier,~V., and Qiao,~R. (2011) A
  “Counter-Charge Layer in Generalized Solvents” Framework for Electrical
  Double Layers in Neat and Hybrid Ionic Liquid Electrolytes. \emph{Physical
  Chemistry Chemical Physics} \emph{13}, 14723--14734\relax
\mciteBstWouldAddEndPuncttrue
\mciteSetBstMidEndSepPunct{\mcitedefaultmidpunct}
{\mcitedefaultendpunct}{\mcitedefaultseppunct}\relax
\EndOfBibitem
\bibitem[Li \latin{et~al.}(2014)Li, Feng, and Cummings]{li2014interfaces}
Li,~S., Feng,~G., and Cummings,~P.~T. (2014) Interfaces of Dicationic Ionic
  Liquids and Graphene: A Molecular Dynamics Simulation Study. \emph{Journal of
  Physics: Condensed Matter} \emph{26}, 284106\relax
\mciteBstWouldAddEndPuncttrue
\mciteSetBstMidEndSepPunct{\mcitedefaultmidpunct}
{\mcitedefaultendpunct}{\mcitedefaultseppunct}\relax
\EndOfBibitem
\bibitem[Merlet \latin{et~al.}(2013)Merlet, Salanne, Rotenberg, and
  Madden]{merlet2013influence}
Merlet,~C., Salanne,~M., Rotenberg,~B., and Madden,~P.~A. (2013) Influence of
  Solvation on the Structural and Capacitive Properties of Electrical Double
  Layer Capacitors. \emph{Electrochimica Acta} \emph{101}, 262--271\relax
\mciteBstWouldAddEndPuncttrue
\mciteSetBstMidEndSepPunct{\mcitedefaultmidpunct}
{\mcitedefaultendpunct}{\mcitedefaultseppunct}\relax
\EndOfBibitem
\bibitem[Limmer \latin{et~al.}(2013)Limmer, Merlet, Salanne, Chandler, Madden,
  Van~Roij, and Rotenberg]{limmer2013charge}
Limmer,~D.~T., Merlet,~C., Salanne,~M., Chandler,~D., Madden,~P.~A.,
  Van~Roij,~R., and Rotenberg,~B. (2013) Charge Fluctuations in Nanoscale
  Capacitors. \emph{Physical review letters} \emph{111}, 106102\relax
\mciteBstWouldAddEndPuncttrue
\mciteSetBstMidEndSepPunct{\mcitedefaultmidpunct}
{\mcitedefaultendpunct}{\mcitedefaultseppunct}\relax
\EndOfBibitem
\bibitem[Roy and Maroncelli(2010)Roy, and Maroncelli]{roy2010improved}
Roy,~D., and Maroncelli,~M. (2010) An Improved Four-Site Ionic Liquid Model.
  \emph{The Journal of Physical Chemistry B} \emph{114}, 12629--12631\relax
\mciteBstWouldAddEndPuncttrue
\mciteSetBstMidEndSepPunct{\mcitedefaultmidpunct}
{\mcitedefaultendpunct}{\mcitedefaultseppunct}\relax
\EndOfBibitem
\bibitem[Edwards \latin{et~al.}(1984)Edwards, Madden, and
  McDonald]{edwards1984computer}
Edwards,~D.~M., Madden,~P.~A., and McDonald,~I.~R. (1984) A Computer Simulation
  Study of the Dielectric Properties of a Model of Methyl Cyanide: I. The Rigid
  Dipole Case. \emph{Molecular physics} \emph{51}, 1141--1161\relax
\mciteBstWouldAddEndPuncttrue
\mciteSetBstMidEndSepPunct{\mcitedefaultmidpunct}
{\mcitedefaultendpunct}{\mcitedefaultseppunct}\relax
\EndOfBibitem
\bibitem[Cole and Klein(1983)Cole, and Klein]{cole1983interaction}
Cole,~M.~W., and Klein,~J.~R. (1983) The Interaction between Noble Gases and
  the Basal Plane Surface of Graphite. \emph{Surface Science} \emph{124},
  547--554\relax
\mciteBstWouldAddEndPuncttrue
\mciteSetBstMidEndSepPunct{\mcitedefaultmidpunct}
{\mcitedefaultendpunct}{\mcitedefaultseppunct}\relax
\EndOfBibitem
\bibitem[Merlet \latin{et~al.}(2012)Merlet, Salanne, and
  Rotenberg]{merlet2012new}
Merlet,~C., Salanne,~M., and Rotenberg,~B. (2012) New Coarse-grained Models of
  Imidazolium Ionic Liquids for Bulk and Interfacial Molecular Simulations.
  \emph{The Journal of Physical Chemistry C} \emph{116}, 7687--7693\relax
\mciteBstWouldAddEndPuncttrue
\mciteSetBstMidEndSepPunct{\mcitedefaultmidpunct}
{\mcitedefaultendpunct}{\mcitedefaultseppunct}\relax
\EndOfBibitem
\bibitem[P{\'e}an \latin{et~al.}(2014)P{\'e}an, Merlet, Rotenberg, Madden,
  Taberna, Daffos, Salanne, and Simon]{pean2014dynamics}
P{\'e}an,~C., Merlet,~C., Rotenberg,~B., Madden,~P.~A., Taberna,~P.-L.,
  Daffos,~B., Salanne,~M., and Simon,~P. (2014) On the Dynamics of Charging in
  Nanoporous Carbon-based Supercapacitors. \emph{ACS nano} \emph{8},
  1576--1583\relax
\mciteBstWouldAddEndPuncttrue
\mciteSetBstMidEndSepPunct{\mcitedefaultmidpunct}
{\mcitedefaultendpunct}{\mcitedefaultseppunct}\relax
\EndOfBibitem
\bibitem[Maroncelli \latin{et~al.}(2012)Maroncelli, Zhang, Liang, Roy, and
  Ernsting]{maroncelli2012measurements}
Maroncelli,~M., Zhang,~X.-X., Liang,~M., Roy,~D., and Ernsting,~N.~P. (2012)
  Measurements of the Complete Solvation Response of Coumarin 153 in Ionic
  Liquids and the Accuracy of Simple Dielectric Continuum Predictions.
  \emph{Faraday discussions} \emph{154}, 409--424\relax
\mciteBstWouldAddEndPuncttrue
\mciteSetBstMidEndSepPunct{\mcitedefaultmidpunct}
{\mcitedefaultendpunct}{\mcitedefaultseppunct}\relax
\EndOfBibitem
\bibitem[Reed \latin{et~al.}(2007)Reed, Lanning, and
  Madden]{reed2007electrochemical}
Reed,~S.~K., Lanning,~O.~J., and Madden,~P.~A. (2007) Electrochemical Interface
  between an Ionic Liquid and a Model Metallic Electrode. \emph{The Journal of
  chemical physics} \emph{126}, 084704\relax
\mciteBstWouldAddEndPuncttrue
\mciteSetBstMidEndSepPunct{\mcitedefaultmidpunct}
{\mcitedefaultendpunct}{\mcitedefaultseppunct}\relax
\EndOfBibitem
\bibitem[Siepmann and Sprik(1995)Siepmann, and Sprik]{siepmann1995influence}
Siepmann,~J.~I., and Sprik,~M. (1995) Influence of Surface Topology and
  Electrostatic Potential on Water/Electrode Systems. \emph{The Journal of
  chemical physics} \emph{102}, 511--524\relax
\mciteBstWouldAddEndPuncttrue
\mciteSetBstMidEndSepPunct{\mcitedefaultmidpunct}
{\mcitedefaultendpunct}{\mcitedefaultseppunct}\relax
\EndOfBibitem
\bibitem[Gingrich and Wilson(2010)Gingrich, and Wilson]{gingrich2010ewald}
Gingrich,~T.~R., and Wilson,~M. (2010) On the Ewald Summation of Gaussian
  Charges for the Simulation of Metallic Surfaces. \emph{Chemical Physics
  Letters} \emph{500}, 178--183\relax
\mciteBstWouldAddEndPuncttrue
\mciteSetBstMidEndSepPunct{\mcitedefaultmidpunct}
{\mcitedefaultendpunct}{\mcitedefaultseppunct}\relax
\EndOfBibitem
\bibitem[Johnson(1928)]{johnson1928thermal}
Johnson,~J.~B. (1928) Thermal Agitation of Electricity in Conductors.
  \emph{Physical review} \emph{32}, 97\relax
\mciteBstWouldAddEndPuncttrue
\mciteSetBstMidEndSepPunct{\mcitedefaultmidpunct}
{\mcitedefaultendpunct}{\mcitedefaultseppunct}\relax
\EndOfBibitem
\bibitem[Nyquist(1928)]{nyquist1928thermal}
Nyquist,~H. (1928) Thermal Agitation of Electric Charge in Conductors.
  \emph{Physical review} \emph{32}, 110\relax
\mciteBstWouldAddEndPuncttrue
\mciteSetBstMidEndSepPunct{\mcitedefaultmidpunct}
{\mcitedefaultendpunct}{\mcitedefaultseppunct}\relax
\EndOfBibitem
\bibitem[Parsons(1990)]{parsons1990electrical}
Parsons,~R. (1990) The Electrical Double Layer: Recent Experimental and
  Theoretical Developments. \emph{Chemical Reviews} \emph{90}, 813--826\relax
\mciteBstWouldAddEndPuncttrue
\mciteSetBstMidEndSepPunct{\mcitedefaultmidpunct}
{\mcitedefaultendpunct}{\mcitedefaultseppunct}\relax
\EndOfBibitem
\bibitem[Stillinger and Lovett(1968)Stillinger, and
  Lovett]{stillinger1968general}
Stillinger,~F.~H., and Lovett,~R. (1968) General Restriction on the
  Distribution of Ions in Electrolytes. \emph{The Journal of Chemical Physics}
  \emph{49}, 1991--1994\relax
\mciteBstWouldAddEndPuncttrue
\mciteSetBstMidEndSepPunct{\mcitedefaultmidpunct}
{\mcitedefaultendpunct}{\mcitedefaultseppunct}\relax
\EndOfBibitem
\bibitem[dos Santos \latin{et~al.}(2008)dos Santos, Müller-Plathe, and
  Weiss]{dos2008consistency}
dos Santos,~D.~J., Müller-Plathe,~F., and Weiss,~V.~C. (2008) Consistency of
  Ion Adsorption and Excess Surface Tension in Molecular Dynamics Simulations
  of Aqueous Salt Solutions. \emph{The Journal of Physical Chemistry C}
  \emph{112}, 19431--19442\relax
\mciteBstWouldAddEndPuncttrue
\mciteSetBstMidEndSepPunct{\mcitedefaultmidpunct}
{\mcitedefaultendpunct}{\mcitedefaultseppunct}\relax
\EndOfBibitem
\bibitem[Otten \latin{et~al.}(2012)Otten, Shaffer, Geissler, and
  Saykally]{otten2012elucidating}
Otten,~D.~E., Shaffer,~P.~R., Geissler,~P.~L., and Saykally,~R.~J. (2012)
  Elucidating the Mechanism of Selective Ion Adsorption to the Liquid Water
  Surface. \emph{Proceedings of the National Academy of Sciences} \emph{109},
  701--705\relax
\mciteBstWouldAddEndPuncttrue
\mciteSetBstMidEndSepPunct{\mcitedefaultmidpunct}
{\mcitedefaultendpunct}{\mcitedefaultseppunct}\relax
\EndOfBibitem
\bibitem[Liu \latin{et~al.}(2014)Liu, Li, Li, Xie, Ni, and Wu]{liu2014strong}
Liu,~X., Li,~H., Li,~R., Xie,~D., Ni,~J., and Wu,~L. (2014) Strong
  Non-classical Induction Forces in Ion-Surface Interactions: General Origin of
  Hofmeister Effects. \emph{Scientific reports} \emph{4}\relax
\mciteBstWouldAddEndPuncttrue
\mciteSetBstMidEndSepPunct{\mcitedefaultmidpunct}
{\mcitedefaultendpunct}{\mcitedefaultseppunct}\relax
\EndOfBibitem
\bibitem[Jungwirth and Tobias(2006)Jungwirth, and
  Tobias]{jungwirth2006specific}
Jungwirth,~P., and Tobias,~D.~J. (2006) Specific Ion Effects at the Air/Water
  Interface. \emph{Chemical reviews} \emph{106}, 1259--1281\relax
\mciteBstWouldAddEndPuncttrue
\mciteSetBstMidEndSepPunct{\mcitedefaultmidpunct}
{\mcitedefaultendpunct}{\mcitedefaultseppunct}\relax
\EndOfBibitem
\bibitem[Feng \latin{et~al.}(2014)Feng, Jiang, Qiao, and
  Kornyshev]{Feng2014Water}
Feng,~G., Jiang,~X., Qiao,~R., and Kornyshev,~A.~A. (2014) Water in Ionic
  Liquids at Electrified Interfaces: The Anatomy of Electrosorption. \emph{ACS
  Nano} \emph{8}, 11685--11694\relax
\mciteBstWouldAddEndPuncttrue
\mciteSetBstMidEndSepPunct{\mcitedefaultmidpunct}
{\mcitedefaultendpunct}{\mcitedefaultseppunct}\relax
\EndOfBibitem
\end{mcitethebibliography}

\newpage
\onecolumngrid
\newpage
\preprint{1}
\setcounter{figure}{0}   
\renewcommand{\thefigure}{S\arabic{figure}}
\begin{center}
\subsection*{\LARGE{Supporting Information}}
\subsection*{\Large{Concentration Fluctuations and Capacitive Response in Dense Ionic Solutions}}
\large{Betul Uralcan, Ilhan A. Aksay, Pablo G. Debenedetti, and David T. Limmer}
\subsection*{\large{Simulation Details}}
\end{center}

In this system, electrolyte species are represented by coarse-grained models where the interactions beyond chemically bonded atoms are represented by Lennard-Jones and Coulombic forces. The force field for [BMIM$^+$][PF6$^-$] is developed by Roy and Maroncelli$^1$ while the model for ACN is taken from Edwards et al..$^2$ ACN and [BMIM$^+$] are represented by three site molecules and [PF6-] is treated as a sphere. The electrodes are modeled as three layers of fixed carbon layers subject to two-dimensional periodic boundary conditions.$^3$ The distance between carbon atoms within each layer is 1.43 \AA, and the distance between layers is 3.38 \AA. The parameters are summarized in TABLE S1. We study fourteen systems, containing $\rho$ = 0.09, 0.12, 0.15, 0.18, 0.23, 0.32, 0.47, 0.54, 0.63, 0.68, 0.77, 0.89, 0.95 and 1 molar fraction of [BMIM$^+$][PF6$^-$].
\vspace{5mm}
%\begin {table}[H]
 %\caption {Table Title} \label{tab:title} 
\begin{center}
\begin{tabular}{ |c|c|c|c|c|c|c|c|c| } 

 \hline
 Site & A & C1 & C2 & C3 & Me & CA & N & CE \\
 \hline
 $\sigma$ (\AA) & 5.06 & 4.38 & 3.41 & 5.04 & 3.60 & 3.40 & 3.30 & 3.37\\
 \hline
 $\epsilon$ (kJ/mol) & 4.71 &	2.56 &	0.36 &	1.83 &	1.59 &	0.42 &	0.42 &	0.23\\
 \hline
 $q ($e) & -0.78	& 0.4374 &	0.1578 	& 0.1848 &	0.269 &	0.129 &	-0.398 &	-\\
 \hline
 $M ($g/mol) & 144.96 &	67.07 &	15.04 &	57.12 &	15.04 &	12.01 &	14.01 &	-\\
 \hline
 $x ($\AA) & 0 &	0 &	0 &	0 &	0 &	0 &	0 &	-\\
 \hline
 $y ($\AA) &0 &	-0.527 &	1.641 &	0.187 &	0 &	0 &	0 &	-\\
 \hline
 $z ($\AA) & 0 &	1.365 &	2.987 &	-2.389 &	0 &	1.46 &	2.63 &	-\\
 \hline
\end{tabular}
\end{center}
%\hspace{0cm}
TABLE S1: Force-field parameters for the molecules of the electrolyte and electrode$^1$$^-$$^3$. A is [PF6$^-$], while C1, C2 and C3 are the three sites of the [BMIM$^+$] cation. Me is the methyl group, and CA and N are the other two sites of acetonitrile. CE is the carbon atoms of the electrode. Crossed parameters are calculated by Lorentz-Berthelot mixing rules.
\vspace{5mm}

The molecular dynamics simulations were conducted in the NVT ensemble using a time step of $1.5$ fs and a langevin thermostat with a time constant of $6$-$9$ ps. Average pressure for all systems is maintained at $10$ atm. For each simulation at $0$ V, $2$-$7$ ns equilibration at $T=400$ K is followed by a $20$-$80$ ns production run from which configurations are sampled every $0.15$ ps.  

\subsection*{\large{Capacitance as a Continuous Function of Electrode Potential}}

In order to determine the evolution of capacitance with applied voltage, we perform additional simulations at seven electrode potentials ($0.15, 0.30, 0.45, 0.60, 0.75, 0.90$ and $1.0$ V). We estimate the probability distribution of the total electrode charge at any potential by combining the data from the simulations performed at various potential differences using the weighted histogram analysis method with the expression

\begin{equation}
- \ln P(Q|0) = - \ln P(Q|\Delta \Psi) + \beta Q \Delta \Psi + \beta {\Delta F} 
\end{equation}
where ${\Delta F} = F(\Delta \Psi) - F(0)$.$^4$

The probability distribution of the total charge $Q$ on the electrodes at $\Delta \Psi = 0$ V is reported in FIG. S1 as a function of $\delta Q / \sqrt{\langle (\delta Q)^2  \rangle}$. In this way, each simulation under an applied potential provides an estimate of the charge distribution at any other potential. Capacitance is then calculated by computing the variance of electrode charge fluctuations from the probability distribution data.
\begin{figure}
\begin{center}
\includegraphics[width=7.5cm]{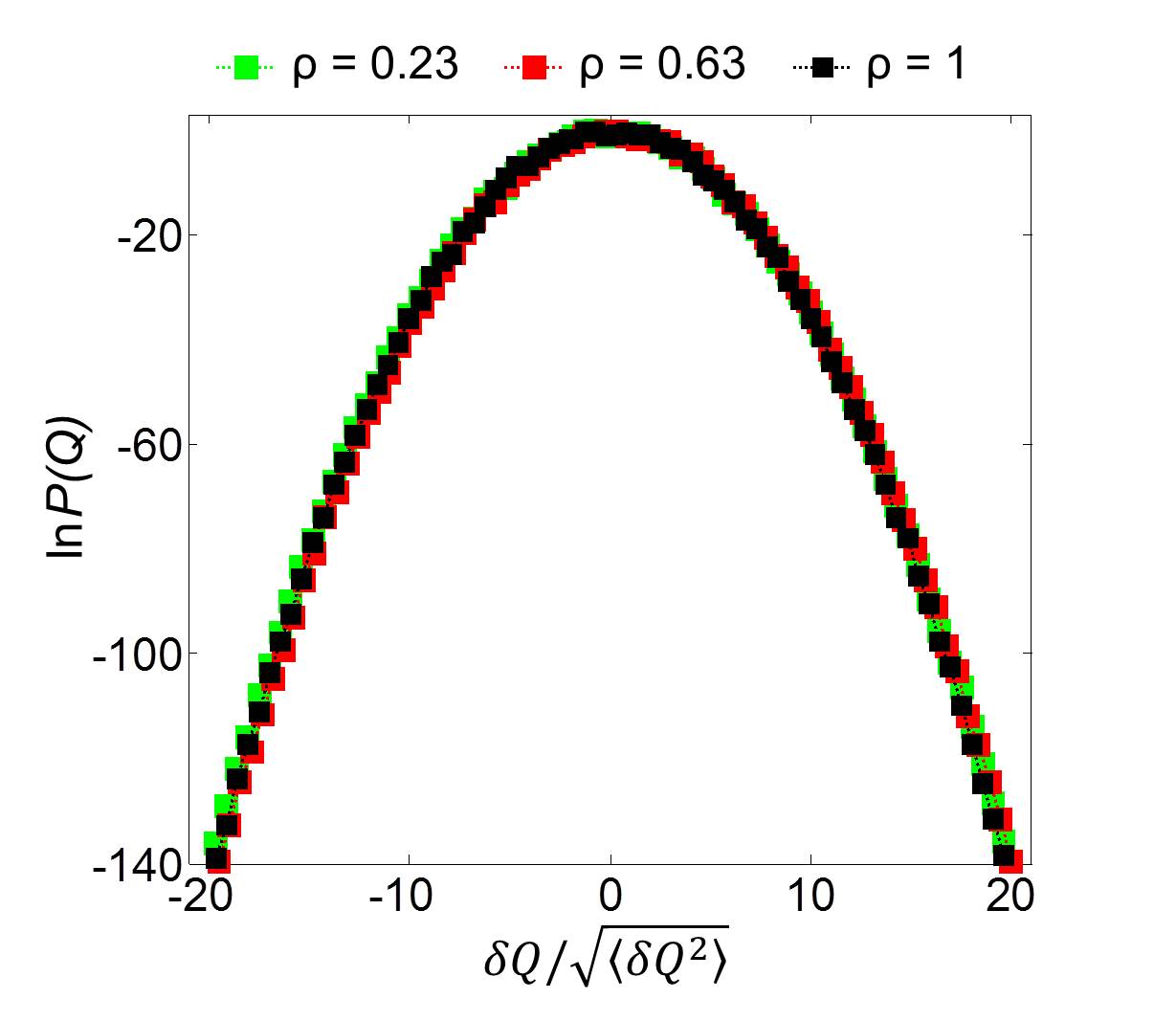}
%\captionsetup{labelformat=empty}
\caption{Probability distribution of the total charge $Q$ on the electrodes at $\Delta \Psi = 0$ V for $\rho$ = 0.23, 0.63 and 1.}
\end{center} 
\end{figure}
\newpage
\subsection*{\large{Bulk Charge Density}}

Radial charge density distributions for bulk electrolyte is computed and fit with a damped harmonic function as in Eq. 5 of the main text. The decay constants $\ell$ obtained from the bulk charge density profiles are given in FIG. S2. The periodicity of oscillations, $q_\mathrm{s}$, is 5.51 \AA. The decay constants obtained from bulk and out-of-plane charge density distributions exhibit similar trends, consistent with the expectation that the electrode interacts with the solutions weakly.
\begin{figure}[h]
\begin{center}
\includegraphics[width=8cm]{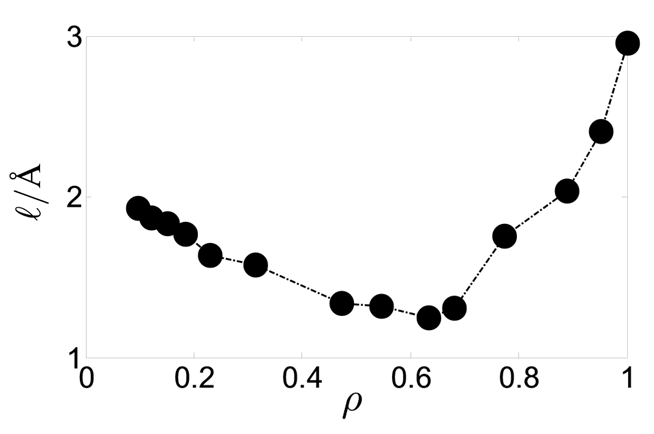}
%\captionsetup{labelformat=empty}
\caption{Dependence of the bulk charge density on concentration. Dependence of the bulk charge density on concentration. Charge density correlation length as a function of ion concentration. See Eq. 5 in the main paper for the definition of correlation length. The black line is a guide to the eye.}
\end{center} 
\end{figure}
%\end {table}
% Produces the bibliography via BibTeX.
\small
\subsection*{\large{References}}
$~^1$ D. Roy and M. Maroncelli, “An improved four-site ionic liquid model,” The journal of physical chemistry letters, 39, 12629-12631 (2010).

$~^2$ D. M. Edwards, P. A. Madden, and I. R. McDonald,  “A computer simulation study of the dielectric properties of a model of methyl cyanide: I. the rigid dipole case,"  Molecular physics 51, 1141-1161 (1984).

$~^3$ M. W. Cole and J. R. Klein, “The interaction between noble gases and the basal plane surface of graphite," Surface Science 124, 547-554 (1983).

$~^4$ D. T. Limmer, C. Merlet, M. Salanne, D. Chandler, P. A. Madden, R. Van Roij, and B. Rotenberg, “Charge fluctuations in nanoscale capacitors," Physical review letters 111, 106102 (2013).

\end{document}